# Data Mining Graphene: Correlative Analysis of Structure and Electronic Degrees of Freedom in Graphenic Monolayers with Defects

Maxim Ziatdinov[1,2], Shintaro Fujii[3], Manabu Kiguchi[3], Toshiaki Enoki[3], Stephen Jesse[1,2], Sergei V. Kalinin[1,2]

[1]Institute for Functional Imaging of Materials, Oak Ridge National Laboratory, Oak Ridge TN 37831, USA

[2]Center for Nanophase Materials Sciences, Oak Ridge National Laboratory, Oak Ridge TN 37831, USA

[3]Department of Chemistry, Tokyo Institute of Technology, Tokyo 152-8551, Japan

To whom correspondence should be addressed:

*E-mail: ziatdinovma@ornl.gov*

*E-mail: sergei2@ornl.gov*

# ABSTRACT

The link between changes in the material crystal structure and its mechanical, electronic, magnetic, and optical functionalities – known as the structure-property relationship – is the cornerstone of the contemporary materials science research. The recent advances in scanning transmission electron and scanning probe microscopies (STEM and SPM) have opened an unprecedented path towards examining the materials structure–property relationships on the single-impurity and atomic-configuration levels. Lacking, however, are the statistics-based approaches for cross-correlation of structure and property variables obtained in different information channels of the STEM and SPM experiments. Here we have designed an approach based on a combination of sliding window fast Fourier transform, Pearson correlation matrix, and linear and kernel canonical correlation, to study a relationship between lattice distortions and electron scattering from SPM data on graphene with defects. Our analysis revealed that the strength of coupling to strain is altered between different scattering channels, which can explain coexistence of several quasiparticle interference patterns in nanoscale regions of interest. In addition, the application of kernel functions allowed us to extract a non-linear component of the relationship between the lattice strain and scattering intensity in graphene. The outlined approach can be further utilized towards analyzing correlations in various multi-modal imaging techniques where the information of interest is spatially distributed and generally has a complex multidimensional nature.

**Introduction.** The central postulation of the contemporary structure-property relation paradigm is that the properties of materials are a direct function of their structure[1-3]. This allows scenarios in which relatively small changes in the material crystal structure may have a decisive impact on the physical properties of the system. For example, introduction of dopant atoms into a lattice of a Mott insulator can turn it into a superconductor whose critical transition temperature ($T_c$) is sensitive to subtle variations of the inter-atomic bond lengths and bond angles in the crystal lattice structure[4-6]. The recently developed framework of materials cartography[3], based on application of graph theory and machine learning algorithms to database of materials ab-initio calculations, allows "breaking" the crystal system down to the individual geometric fragments and identifying the geometric features that most influence the physical property of interest (for example, $T_c$).

Despite achieving substantial progress in theoretical prediction of structure-property relationships, the experimental identification of a connection between material functionalities and variations in the crystal lattice geometry has been typically difficult and indirect. Recently, the progress in the development of high-resolution, multi-modal scanning transmission electron and scanning probe microscopies (STEM and SPM) has allowed researchers performing simultaneous measurements of materials structural parameters (e.g., lattice distortion) and functional properties (e.g., perturbation in local density of states) in real space with sub-nanometer precision[2,7,8]. This potentially allows studying the structure–property relationship in materials on the single-defect and atomic-configuration levels. However, methods to cross-correlate information obtained in different information channels and to describe the obtained correlation in terms of certain (linear or non-linear) physical models are very limited and yet are increasingly necessary due to the ever-growing volumes of relevant experimental data.

Here we illustrate how to use information on local atomic coordinates and parameters of quasiparticle scattering obtained from atomically-resolved SPM images of graphene samples with defects to mine structure-property relationships in these systems. Due to unique properties of the (undistorted) graphene, in which the charge density oscillations are commensurate with the underlying atomic lattice[9-11], both structure and property information channels can be accessed in a single SPM image acquired in the electron tunneling regime[10,11]. By applying a combination of the sliding window fast Fourier transform, Pearson correlation matrix, and linear and kernel canonical correlation analysis, we showed the presence of a non-negligible correlation between the nanoscale lattice strain and the intensities of inter-valley quasiparticle scattering at defect sites. Our analysis also revealed that the strength of coupling to strain is altered between different scattering channels, which can explain an emergence of more than one quasiparticle scattering pattern in the sample (the so-called fine structure of the electronic superlattice). Finally, application of the kernelised canonical correlation analysis aided in extracting a non-linear component of the relationship between the lattice strain and scattering intensity in graphene.

**Experimental systems.** To illustrate our approach we have selected two different graphenic systems. The first system is a monolayer of reduced graphene oxide on Au(111) substrate, which we denote as $G_O$. Figure 1a shows the representative current map of such system obtained in the small bias regime (2 mV) of a conductive atomic force microscopy (c-AFM) where the tunneling current signal is proportional to the local density of states at the Fermi level[12]. The second system is the top graphene layer of graphite, peppered with the hydrogen-passivated vacancies and nanosized holes, $G_H$[13]. The spatial distribution of π-electronic states in this system, obtained *via* scanning tuneling microscopy, is shown in Fig. 1b. More experimental details on both systems, including sample preparation procedures, can be found in Ref. 12[12] and Ref. 13[13]. The global 2-dimensional fast Fourier transform (2D FFT) performed on the data from Fig. 1a and 1b shows a very similar reciprocal space patterns for both systems which are characterized by a presence of two well-defined hexagons (Fig. 1c and 1d). These two hexagons are rotated by 30º with respect to each other, and their lattice constants differ by a factor of $\approx \sqrt{3}$. The outer hexagon corresponds to graphene reciprocal lattice points. The deformation of the perfect graphene lattice would result in a shift of the centers of the FFT $k_l = (k_1, k_2, k_3)$ spots from their original high-symmetry positions depicted in Fig. 1e. The inner hexagon corresponds to a formation of the electronic superlattice with an average unit cell of $(\sqrt{3} \times \sqrt{3})R30º$ where $R = a_0$ is a translational vector of unperturbed graphene lattice (hereafter referred as *R*3 superlattice). We denote the FFT spots corresponding to *R*3 superstructure as $K_e = (K_1, K_2, K_3)$. Due to the symmetry properties of the FFT method, it is sufficient to analyze the correlations associated with only 3 distinct FFT peaks (not paired by the inversion symmetry) in each hexagon.

The formation of the R3 superlattice in undistorted graphene can be explained as the constructive interference between incident and (back)scattered states associated with electron valleys at the opposite corner points, $K^+$ and $K^-$, of the hexagonal Brillouin zone (BZ)[10,11,14]. The corresponding momentum scattering is $\hbar q = \hbar(k_F + G)$, where $k_F$ and $G$ are Fermi wavevector and the reciprocal lattice vector of graphene, respectively see Fig. 1 (e). Owing to the symmetry of graphene lattice, there are three backscattering channels that we denote as $K_i^+ \rightarrow K_i^-$ ($i$=1, 2, 3). The circle-like shapes of the FFT spots associated with backscattering channels (reciprocal lattice) in Fig. 1c and Fig. 1d are due to the distribution of scattering (lattice) vectors across the experimental image. Note that for point defects that preserve the lattice symmetry, the scattering probability is equivalent in all the three channels resulting in the same intensity of the six inner hexagon's peaks in FFT. However, the equivalency between different scattering channels may not necessarily hold for the distorted graphene lattice. Indeed, at the local scale we are able to observe a strong modulation in the relative intensities of the inner hexagon FFT spots that results in a fine structure of the electronic superlattice around the defects in the real space (see Fig. 1b, 1f and 1g). The precise origin of such a fine structure is not exactly understood at the present moment.

**Sliding FFT.** Our goal is to analyze a structure-property relationship in the two graphenic systems by studying the correlation between local lattice distortions associated with $k_l$ peaks and electronic features associated with $K_e$ peaks (see Fig. 1e). In the first step, we shift a square window of size ($w_x$, $w_y$) across the input image ($T_x$, $T_y$) in series of steps $x_s$ and $y_s$ such that the entire image is scanned. At each step, the 2D FFT is computed for the image portion that lies within the window. This procedure is known as sliding FFT method and its application to scanning probe microscopic data has been described in detail in our earlier publication[15]. The amplitudes and coordinates of the selected peaks are extracted from each 2D FFT image by fitting them with 2D Gaussian distribution,

$$G(Q_x, Q_y) = A \exp\left[-\left(\frac{(Q_x-Q_x^0)^2+(Q_y-Q_y^0)^2}{2\sigma^2}\right)\right], \quad (1)$$

where $A$ is the peak amplitude, $(Q_x, Q_y)$ are the Cartesian coordinates of the peak position, and $\sigma$ is the standard deviation.

We first compute the lattice strain $\varepsilon$ defined as the variation of the lattice vector along the three directions shown in Fig. 2a,

$$\varepsilon_i = (a_i - \bar{a}_i)/\bar{a}_i. \tag{2}$$

Here $a_i$ is the unit cell vector calculated for each pixel of the resultant sliding FFT maps using a standard relation between real space and reciprocal space lattices in graphene (index $i$ stands for three different directions) and $\bar{a}_i$ is the mean value of unit cell vector in the overall image; for the randomly fluctuating strain fields the mean value is close to that in the unperturbed lattice, $\bar{a}_i \approx a_0$. We also compute the variations in the unit cell area, $\Delta S = (s - \bar{s})/\bar{s}$, where $s$ is the mean of the 3-element vector such that $s_{ij} = 1/2(a_i a_j)$. In a similar fashion, the relative shift $\Delta K$ in the positions of the $K_e$ vectors of the $R3$ superlattice is determined as $\Delta K_i = (K_i - \bar{K}_i)/\bar{K}_i$. The scattering intensities $I_{K_i} = I(K_i^+ \to K_i^-)$ associated with the $R3$ superlattice are given by the amplitudes of the corresponding $K_e$ spots extracted directly from the sliding FFT maps.

**Mining the linear relationships.** We start with exploring a relationship between the lattice strain and parameters of the $R3$ electronic superlattice in graphene via pairwise correlation matrix analysis. The correlation parameter for each pair of variables $x$ and $y$ is defined as a linear Pearson correlation coefficient,

$$r_{xy} = \frac{\sum_{i=1}^{N}(x_i - \bar{x})(y_i - \bar{y})}{\sqrt{\sum_{i=1}^{N}(x_i - \bar{x})^2}\sqrt{\sum_{i=1}^{N}(y_i - \bar{y})^2}} \tag{3}$$

where $\bar{x}$ is the mean of $x$, $\bar{y}$ is the mean of $\bar{y}$, and $N$ is a number of scalar observations. The results for $G_O$ and $G_H$ samples are shown in Fig. 2b and 2c respectively. We note in passing that the typical range of the lattice strain values in (2) was within ±0.1 (±10%) in both samples. Remarkably, the dependence of the scattering intensity on the unit cell area and corresponding strain components shows quite an opposite behavior for our two samples. In the $G_O$ sample, there is a strong negative correlation (maximum absolute value $|r|_{max}$~0.51) for all backscattering channels (Fig. 2b). This is not the case, however, for $G_H$ sample (Fig. 2c), in which the scattering intensities associated with two channels ($I_{K_1}$ and $I_{K_2}$) show a considerable positive correlation ($|r|_{max}$~0.59). Possible origin of different behavior of coupling coefficients in two samples will be explained below.

We first analyze the correlation matrix results for $G_O$ sample. We recall that some oxygen functional groups, and in particular epoxy group, are known to remain on the surface of reduced graphene oxide.[16,17] The effect of the epoxy functional groups on the structure of graphene monolayer is two-fold. First, they induce the out-of-plane distortions ("rippling") of the graphene sheet.[17-19] As the experimental image gives a 2D-projection of a 3D lattice structure, the rippling of graphene sheet tilts the lattice with respect to the viewing direction causing an apparent contraction of the lattice constant. Second, the presence of epoxy groups is known to induce changes in the C-C bond lengths, including a considerable stretching of the nearby bonds up to 1.58 Å[18,20,21], that may well be hidden from our view "under" the curvature regions in the experimental image. Both curvature[22] and stretched C-C bonds[23,24] can lead to the enhancement of the density of states in the surrounding region, although a simple correlation analysis may not allow us to disentangle these two contributions. Particularly, as graphene bonds are stretched, first-principles calculations show emergence of new peaks in the density of states that shift towards the Fermi level with increasing the length of C-C bond[24]. The dependence of the density of states per area on the bond length can also be understood from the simple tight-binding (TB) model perspective. In the nearest-neighbor TB approximation, the density of states $D(E)$ in monolayer graphene is given by[25]

$$D(E) = \frac{|E|}{\pi\sqrt{3}\gamma^2}, \qquad (4)$$

where $\gamma$ is the nearest neighbor hopping parameter. The dependence of the hopping parameter $\gamma$ on the bond length can be described in terms of the exponential decay model,[26]

$$\gamma \cong \gamma_0 \exp(-\tau\varepsilon), \tag{5}$$

where $\tau$ is typically assigned values between 3 and 4.[27] It follows from (4) and (5) that in agreement with first-principles calculations, the expanded lattice constant in the curved regions can lead to a higher density of states. The enhancement of density of states can in turn increase the number of states participating in the scattering between different valleys at the given energy "under" the curved regions of $G_O$ sample. This can explain the apparent negative correlation coefficient between strain components and the scattering amplitude in Fig. 2b.

We now turn to the $G_H$ sample. As the surface of this sample is peppered with mono- and multi-vacancies,[13] as well as with larger holes,[28] we expect an alteration of the in-plane bond distances and variations in associated nanoscale strain fields across the surface. It should be noted that in contrast with the $G_O$ sample, the out-of-the-plane distortions in the $G_H$ sample are mainly limited to the immediate periphery of the monovacancy-hydrogen complexes and that no hydrogen atoms are adsorbed on the surface in other regions.[13] Thus, the mainly positive correlation between the strain components and the scattering amplitude seen in Fig. 2c shows that in agreement with equations (4) and (5) the density of states available for scattering is enhanced with increasing bond length. Notably, the character of coupling between the strain and the scattering intensity in the $G_H$ sample is clearly altered for $K_2^+ \to K_2^-$ backscattering channel as compared to $K_1^+ \to K_1^-$ and $K_3^+ \to K_3^-$ channels which can be easily seen from the correlation matrix in Fig. 2c. Indeed, the $I_{K_2}$ intensity component shows significantly lower positive correlation values for $\varepsilon_1$ and $\varepsilon_3$ components of strain, as well a negative value for $\varepsilon_2$ strain component. As shown in Fig. 1f and 1g, a suppression of the contribution from FFT $K_2$ spots is associated with the formation of the staggered dimer-like superlattice in the real space image, whereas all the three $K_e$ spots are required to produce the hexagonal superlattice. The fine structure of the electronic superlattice associated with the coexistence of staggered dimer (sometimes characterized as "honeycomb") and hexagonal superlattices has been consistently observed in different STM experiments on graphenic surfaces.[14,29,30] Our correlation matrix analysis suggests that a different coupling of backscattering channels to strain fields can explain this fine structure.

Finally, we note that in both systems we found only a relatively small correlation ($|r|_{max}$~0.33) between the lattice strain and the variation in the momentum space positions of the electron scattering maxima. We recall that in the undeformed graphene lattice the $K_e$ scattering maxima are located at the corners of the graphene BZ (see Fig. 1e) and that the coordinates of the latter in both deformed and undeformed lattices are obtained through a direct linear transformation of the reciprocal space lattice vectors. Hence, our correlation analysis indicates that the location of the electron scattering maxima ("Dirac valleys") in the deformed graphene lattice do not coincide with the position of the BZ corners.

The standard Pearson correlation analysis that was used so far is limited to analyzing bivariate correlations. Canonical correlation analysis (CCA), on the other hand, allows grouping the variables in each multivariate dataset such that optimal correlation is achieved between two sets[31]. Specifically, CCA solves the problem of finding basis vectors **w** and **v** for two multi-dimensional datasets $X$ and $Y$ such that the correlation between their projections $\mathbf{x} \to \langle \mathbf{w}, \mathbf{x} \rangle$ and $\mathbf{y} \to \langle \mathbf{v}, \mathbf{y} \rangle$ onto these basis vectors is maximized (see schematics in Fig. 3a). The canonical correlation coefficient $\rho$ can be expressed as[32]

$$\rho = \max_{\mathbf{w},\mathbf{v}} \frac{\mathbf{w}' C_{xy} \mathbf{v}}{\sqrt{\mathbf{w}' C_{xx} \mathbf{w} \mathbf{v}' C_{yy} \mathbf{v}}}, \tag{6}$$

where $C_{xx}$, $C_{yy}$ are auto-covariance matrices, and $C_{xy}$, $C_{yx}$ are cross-covariance matrices of **x** and **y**. The projections $\mathbf{a} = \mathbf{w}'\mathbf{x}$ and $\mathbf{b} = \mathbf{v}'\mathbf{y}$ represent the first pair of canonical variates.

We apply the CCA method to analyze the dependency of the scattering intensities on variations of the lattice parameter. The canonical correlation between the components of the *apparent* strain and the intensity of electron scattering in $G_O$ sample has a value of $\rho = 0.50$. The associated canonical variables scores are

$$a_i^{strain} = 0.31(\varepsilon_1)_i + 0.73(\varepsilon_2)_i + 0.32(\varepsilon_3)_i$$
$$b_i^{ampl} = -0.37(I_{K1})_i - 0.41(I_{K2})_i - 0.80(I_{K3})_i \tag{7}$$

where the magnitudes of the coefficients before the variables give the optimal contributions of the individual variables to the corresponding canonical variate. For $G_H$ samples, the CCA correlation analysis returns a value of $\rho = 0.62$ and the canonical scores are

$$a_i^{strain} = 0.37(\varepsilon_1)_i + 0.50(\varepsilon_2)_i + 0.36(\varepsilon_3)_i$$
$$b_i^{ampl} = 0.39(I_{K1})_i - 0.33(I_{K2})_i + 0.80(I_{K3})_i \tag{8}$$

Note that for the $G_H$ sample in (8) the coefficient before $I_{K_2}$ has a negative value, whereas the rest of the coefficients are positive. This shows that the scattering intensity associated with the $K_2^+ \to K_2^-$ channel responses to the changes in the bond lengths in the opposite fashion as compared to $K_1^+ \to K_1^-$ and $K_3^+ \to K_3^-$ channels. This distinctive coupling of the strain to the $K_2^+ \to K_2^-$ scattering channel is in the agreement with Pearson correlation matrix analysis in Fig. 2c and confirms that the origin of superlattice fine structure in $G_H$ sample stems from a subtle connection between certain combination of strain fields to the scattering intensity in different channels. Likewise, for the $G_O$ sample in (7), the scattering intensity associated with $K_3^+ \to K_3^-$ backscattering channel and the $\varepsilon_2$ component of strain show the strongest contribution to their respective canonical variates, hinting at the important role of the non-uniform strain-scattering relation in occurrence of various superlattice patterns observed in Fig. 1a. Note that for both $G_O$ and $G_H$ cases the canonical correlation plots in Fig. 3b and Fig. 3c clearly show a presence of nonlinear associations between strain and scattering intensities. In the remaining part of the paper, we will discuss how such a non-linear features can be discovered in the combinatorial library of distortion modes and electronic scattering.

**Mining the non-linear associations.** Due to its linearity, the CCA may not extract useful descriptors from the data in the presence of non-linear relationships between variables. The kernelization of CCA offers an alternate solution by first performing a non-linear mapping $\varphi_x$ and $\varphi_y$ of the data points to a higher dimensional space,

$$\varphi_x: \mathbf{x} \to \varphi_x(\mathbf{x}) \in \Phi_x$$
$$\varphi_y: \mathbf{y} \to \varphi_y(\mathbf{y}) \in \Phi_y, \tag{9}$$

and then performing CCA in the new feature space. In such cases, the linear model found in the new feature space corresponds to a nonlinear model in the input space[32]. The computational complexity arising from the high dimensionality mapping is addressed by using the kernel functions,[32-35]

$$k_x(\mathbf{x}_i, \mathbf{x}_j) = \langle \varphi_x(\mathbf{x}_i), \varphi_x(\mathbf{x}_j) \rangle$$
$$k_y(\mathbf{y}_i, \mathbf{y}_j) = \langle \varphi_y(\mathbf{y}_i), \varphi_y(\mathbf{y}_j) \rangle. \tag{10}$$

We look for a correlation between functions $f \in \Phi_x$ and $g \in \Phi_y$ of the form,

$$f(x) = \sum_{i=1}^{N} \alpha_i k_x(\mathbf{x}_i, \mathbf{x})$$
$$g(y) = \sum_{i=1}^{N} \beta_i k_y(\mathbf{y}_i, \mathbf{y}), \tag{11}$$

where $\alpha$ and $\beta$ are the expansion coefficients. Then, similar to linear CCA, the objective of kernel CCA can be written as[32,33]

$$\rho = \max_{\alpha, \beta} \frac{\alpha' K_x K_y \beta}{\sqrt{\alpha' K_x^2 \alpha \cdot \beta' K_y^2 \beta}}, \tag{12}$$

where $(K_x)_{ij} = k_x(\mathbf{x}_i, \mathbf{x}_j)$ and $(K_y)_{ij} = k_y(\mathbf{y}_i, \mathbf{y}_j)$ are Kernel matrices (see Fig. 4a for schematics). The regularization parameter is usually introduced to kernel CCA model for avoiding overfitting problems in a higher dimensional space.[32]

We use the polynomial kernel (polykernel) functions of the form $k(\mathbf{a}, \mathbf{b}) = (\langle \mathbf{a}, \mathbf{b} \rangle + \eta)^d$ which are well suited for extracting non-linear relationships that may occur in our system. The corresponding kernel Hilbert space is a finite-dimensional vector space, whose dimension depends only on $d$. The first degree polykernel, $d = 1$, captures linear correlations, and is equivalent to linear CCA. Second degree polykernel, $d = 2$, captures all second order statistics, and so forth. We started with testing kernel CCA on synthetic datasets with strong non-linear relationships between stimulus and response. The dependency of the correlation coefficient on the polynomial order of kernel function for several synthetic datasets is shown in Fig. 4c. One can immediately see that the quadratic dependency between stimulus and response can easily be captured by the $d = 2$ polykernels. On the other hand, capturing exponential relation requires polykernels of the order as high as $d = 5$. Overall, observing the behavior of the kernel correlation coefficient with increasing the dimension of the corresponding kernel Hilbert space can provide an important information about the type of non-linear relationship between the multivariate datasets under the consideration. We also demonstrated in Fig. 4c that the application of the polynomial kernels to the synthetic periodic function returns a very low correlation value, indicating that in general the appropriate kernel function must be selected for each physical problem. It is noteworthy that a presence of "distortion" within stimulus-response relationship associated with a higher order correlation between some of the variables from two multidimensional datasets decreases the overall correlation between corresponding canonical variates for a given polykernel. The overall canonical correlation values can also be reduced due to a substantial level of noise in some variables (for real-world physical systems this may happen due to specific limitations on measuring accurately certain components of forces and fields).

We now describe an application of the kernel CCA to finding a non-linear relationship between lattice strain components and scattering intensity in $G_O$ and $G_H$ samples. As discussed earlier in the paper, we expect a presence of the non-linear contribution to the density of states (and associated scattering intensity) in a form of $D(E) \sim \exp(2\tau\varepsilon)$. Figure 4d shows the values of kernel correlation coefficient for different polykernel orders for the $G_O$ sample. The $d = 1$ case reproduces perfectly the results of linear CCA. Overall, the dependency of the correlation coefficient on polykernel order closely resembles the test result for exponential relation between stimulus and response (see Fig. 4c). This implies that, using kernel CCA, we were able to uncover the "hidden-under-the-curvature" component of the correlation between the modulation of the lattice constant and the scattering intensity in the $G_O$ sample. Remarkably, we found very similar behavior of the kernel correlation coefficient in the $G_H$ sample that does not have significant out-of-the-plane curvature. This further confirms that the kernel CCA allowed us uncovering non-linear changes in the scattering intensities originating from the altered C-C distances for both samples. The overall lower values of canonical correlation for $G_O$ sample are likely due to the somewhat "noisier" datasets of strain and intensity maps in this sample. It should be noted that larger values of $\tau$ (up to $\tau \sim 6$) and/or strain (up to 40%) are in general required to get a dominant non-linear contribution from the exponential term; otherwise the relationship can be well approximated by a linear CCA ($d = 1$). The discrepancy may originate from not taking into account in our simplistic model an effect of the underlying substrate on the resultant charge density distribution as well as from neglecting contributions from hopping parameters beyond the nearest neighbor. In addition, certain variations in the bond length may produce a bond-centered charge order that has the same symmetry as the scattering-produced $R3$ superlattice, thus enhancing the signal of interest.[36]

**Conclusions.** In conclusion, we have designed a procedure for data mining structure-property relations from atomically-resolved images of graphene. By applying a combination of the sliding FFT, Pearson correlation matrix, and linear and kernel canonical correlation analysis, we were able to extract detailed information on the correlation between the nanoscale strain and the parameters associated with electron scattering. We found that the expansion of the lattice constant results in the enhanced scattering intensity which was attributed to the increased density of states available for scattering in the stretched areas of the graphene lattice. Our analysis also revealed that the strength of coupling to strain is altered between different scattering channels which can explain an emergence of more than one superperiodic patterns in the samples (the so-called fine structure of the superlattice). Finally, using a kernelized version of canonical correlation analysis, we uncovered the presence of non-linear associations between the strain components and the intensity of electron scattering in graphene. In future, the approach outlined here can be applied to scanning probe datasets representing a combination of atomically-resolved topographic imaging and spectroscopic grids obtained on the same area. This would allow finding, for example, a correlation between the subtle variations of strain and the magnitude of a gap in the superconducting materials. By the same token, our approach can be extended to correlative analysis of the combined electron energy loss spectroscopy and STEM (the so-called STEM-EELS[37]) experimental data, in which the information on the position, electronic structure, and chemical bonding of different atomic columns can be acquired through different channels.

**Acknowledgements:**

This research was sponsored by the Division of Materials Sciences and Engineering, Office of Science, Basic Energy Sciences, US Department of Energy (M.Z. and S.V.K.). Research was conducted at the Center for Nanophase Materials Sciences, which also provided support (S.J.) and is a DOE Office of Science User Facility. S.F., M.K., and T.E. acknowledge support from Grants-in-Aid for Scientific Research (No. 20001006, No. 23750150, and No. 25790002) from the Ministry of Education, Culture, Sports, Science and Technology of Japan. M.Z. thanks Rama K. Vasudevan (ORNL) for proofreading the manuscript.

**Methods:**

The $G_O$ sample was prepared by chemical and thermal reduction of single-layer oxidized graphene on Au(111). The $G_H$ sample was prepared by sputtering the topmost graphene layer of graphite with low-energy $Ar^+$ ions followed by exposure to atomic hydrogen environment and annealing. For more details of the samples preparation procedure, see Ref. 12 and Ref. 13. The data analysis was carried out using Matlab. For sliding FFT procedure, the size of the input image was (874 px × 874 px) for $G_O$ and (768 px × 768 px) for $G_H$. The size of the sliding window was set at $w_x = w_y = 256$ px and the step of the window shift was set at $x_s = y_s = 16$ px. Mean-centering and normalization of the data was performed prior to application of linear and kernel CCA.

# FIGURES

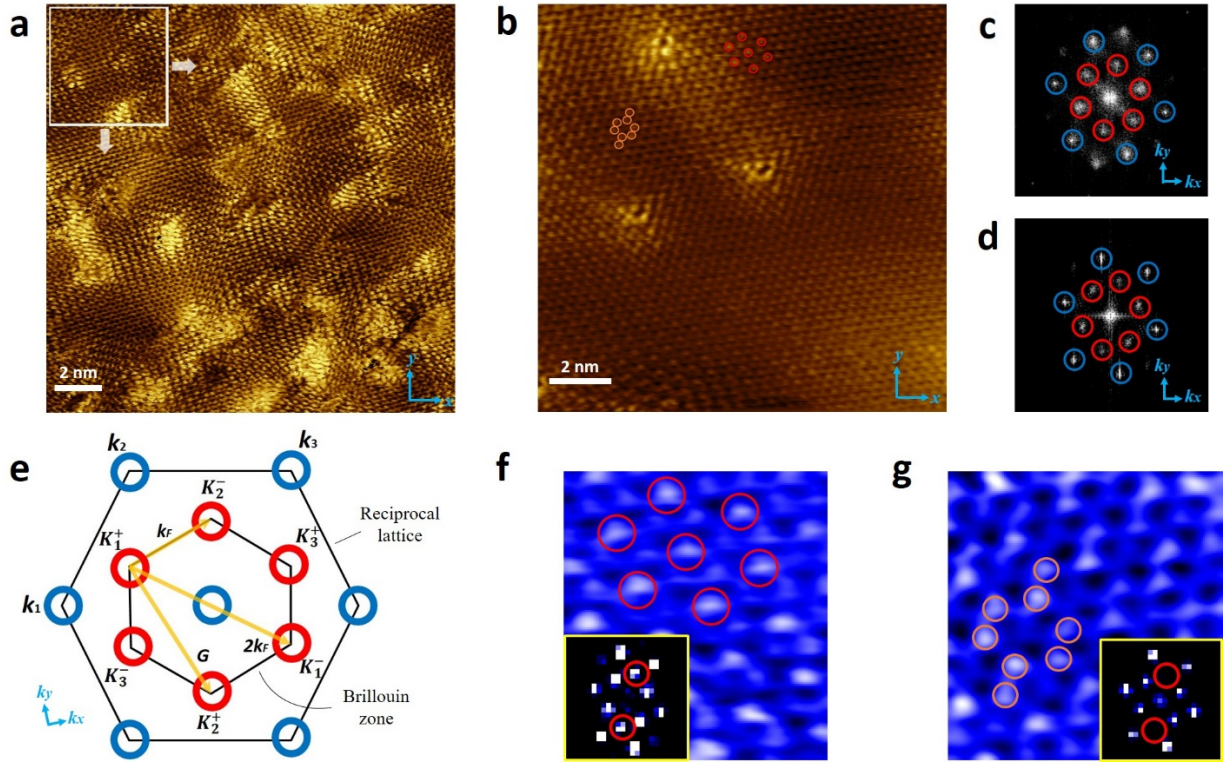

**FIGURE 1. Imaging π electronic states in two different graphenic samples.** (a) Low-bias ($U_s$=2 mV) current-mapping c-AFM image of reduced graphene oxide on Au(111) substrate. The sliding window used for our analysis is schematically shown in the top left corner. (b) STM image of the top graphene layer of graphite with vacancy-hydrogen complexes ($U_s$=100 mV, $I_{setpoint}$=0.9 nA). The 2D FFT transform of data in (a) is shown in (c) and the 2D FFT of data in (b) is shown in (d). (e) Schematics of graphene reciprocal space structure explaining pattern in (c) and (d). (f) Hexagonal superperiodic lattice and its 2D FFT. (g) Staggered-dimer-like electronic superlattice and its 2D FFT. Both superlattices are also marked in (b).

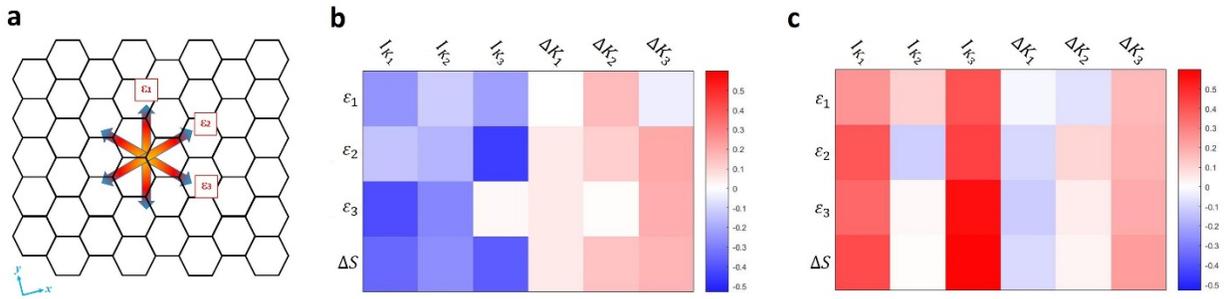

**FIGURE 2. Correlation matrix analysis.** (a) Schematic depiction of 3 different strain components used in our analysis. (b, c) Pairwise Pearson correlation matrix for $G_O$ sample (b) and for $G_H$ sample (c). See text for further details.

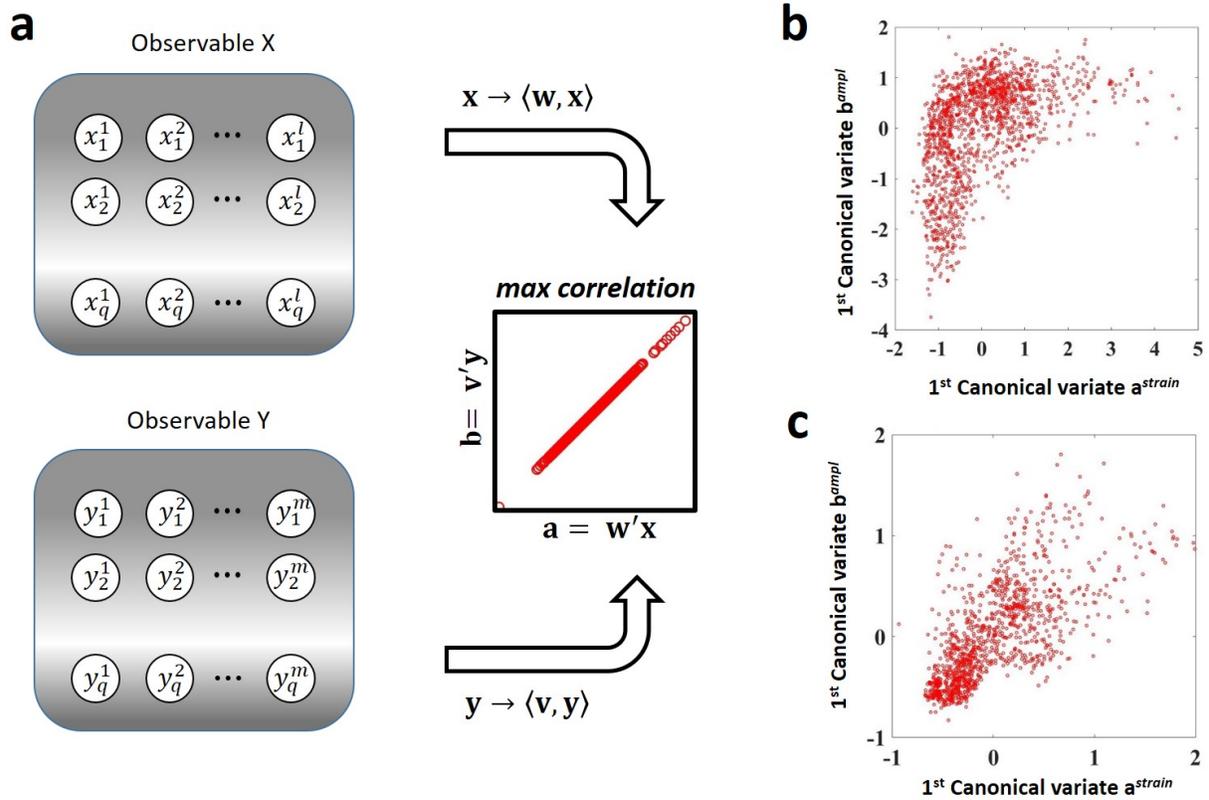

**FIGURE 3. Canonical correlation analysis**. (a) Schematics of CCA workflow. (b) Plot of the canonical variable scores for the correlation between strain components and scattering intensity for the $G_O$ sample. (c) Same for the $G_H$ sample.

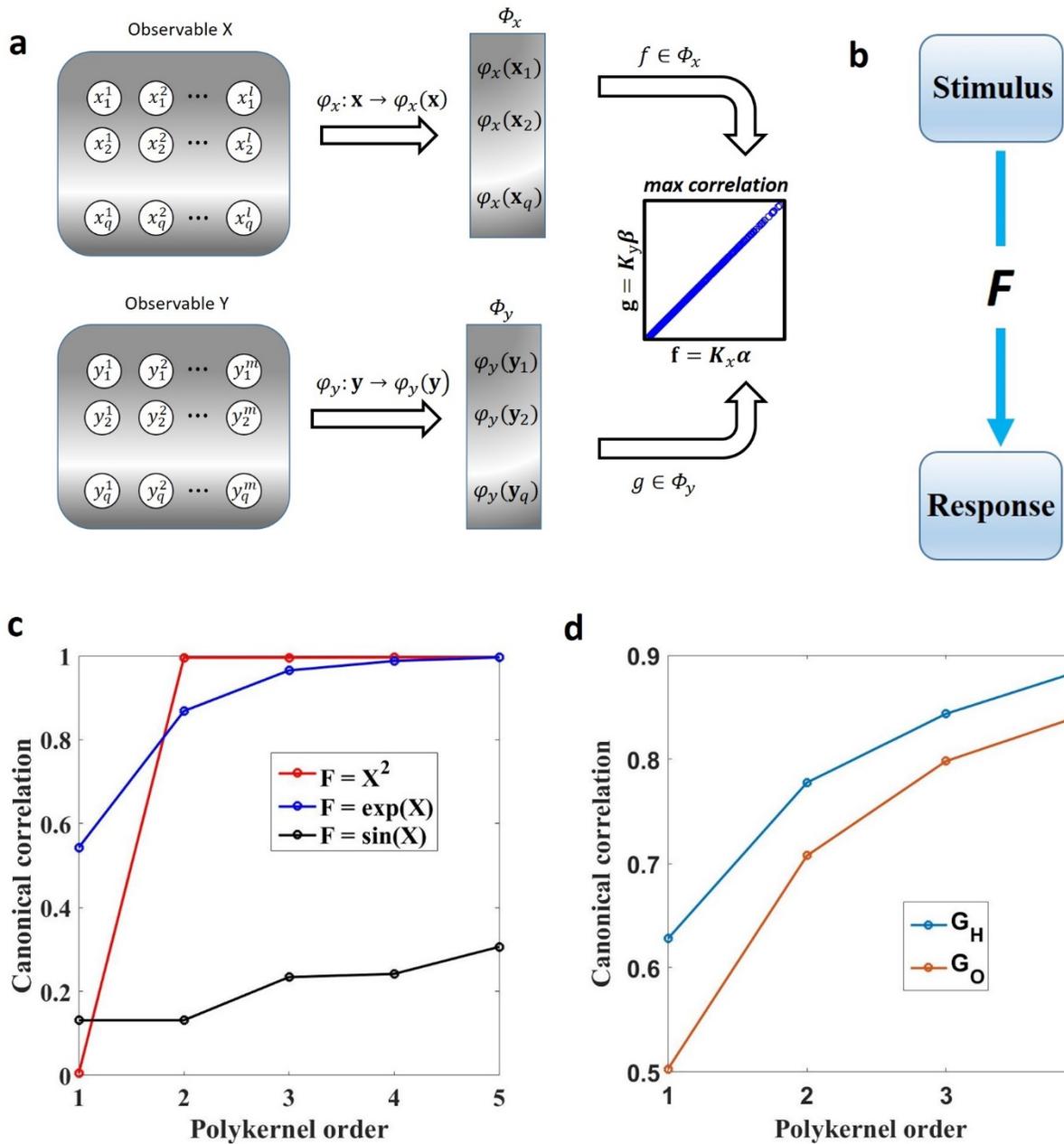

**FIGURE 4. Kernel canonical correlation analysis**. (a) Schematics of kernel CCA workflow. (b) Stimulus-response relationship is defined by function $F$; the purpose of the current analysis is to uncover the function $F$ for the multivariate datasets under consideration. (c) Dependence of the kernel CCA correlation coefficient on the order of kernel polynomial for different synthetic datasets. (d) Application of polynomial kernel to the experimental dataset on $G_O$ and $G_H$.